\documentstyle[aps,epsfig]{revtex}
\begin{document}
\draft
\title{Calculation of Dip Features in
the SIS Conductance of D-Wave Superconductors
Using Eliashberg Formalism }

\author{L. Coffey}

\address{Illinois Institute of Technology,\\
Chicago, Illinois 60616}

\date{\today}
\maketitle

\begin{abstract}
Recent SIS tunneling conductance measurements on a range of Bi2212 crystals
with varying oxygen doping have revealed new information on the behavior of a dip
feature in the tunneling conductance. Calculations presented here show that the
observed variation in position of the dip can be generated in an Eliashberg formalism
for a d-wave superconducting state based on a single peak spectral weight function. 
\end{abstract}

\section{Introduction}

Recent measurements \cite{JFZ}  of the SIS tunneling conductance of a series of Bi-2212 
superconducting crystals, with oxygen doping ranging from overdoped to underdoped, have revealed 
a variation with doping in the position of a dip relative to the main tunneling conductance peak. 
The separation between the main conductance peak and the dip, 
expressed in meV \cite{JFZ}, agrees closely with neutron scattering 
measurements of the energy of a magnetic resonance mode in these superconductors.

Theoretical work on the origin of dip features in tunneling and
photoemission in the cuprates has been in existence for several years \cite{Coffey}.
An example of more recent calculations is that of \cite{Manske} which, using a strong coupling method, 
explored the role of spin fluctuations in producing a d-wave symmetry superconducting
state. In that work, which  shows a dip in the calculated density of states,
the experimentally measured resonance mode was associated 
with a peak in the calculated spin fluctuation spectrum near the Q vector $(\pi \;, \pi$).
The recent experimental observations, that the energy separation between the main
SIS conductance peak and the dip minimum agrees closely with the measured 
mode energy in Bi-2212 \cite{JFZ}, present a good test of strong coupling scenarios
based on a pairing mechanism arising from spin fluctuations.
 
As a reminder of what to look for in the density of states, consider 
a strong coupling Eliashberg calculation for an s-wave
symmetry superconductor with a spectral weight, $\alpha^{2}F(E)$, consisting
of a single peak at an energy $E_{mode} = 1.0$ as depicted in figure (1).
This will lead to a dip feature in the
resulting superconducting density of states. The dip has an onset at
an energy $E=\Delta \; + E_{mode}$, with the position of the dip minimum occuring at
a higher energy \cite{Schrieffer} as illustrated in figure (2). For a coupling strength 
$\lambda \; = \; 2 \;\int \alpha^{2}F(E)/E$ of the order of  2.0, for example,
the position of the dip minimum is located noticeably further out from the onset 
threshold of the dip in the density of states. Consequently, for the s-wave strong coupling case, 
the energy separation between the exact position of the dip minimum and the main
density of states peak will not be an accurate measure of the energy of
the original peak in the underlying $\alpha^{2}F(E)$. This is the case whether
the density of states is studied, or the corresponding SIS conductance curves
which result from a convolution of two such s-wave density of states. 

The aim of calculations presented here, which are based on conventional
Eliashberg strong coupling theory, is to study variations in the
position of dip features in the density of states resulting from a model 
spectral weight with single peak at $E=E_{mode}$
for a  d-wave symmetry superconductor.
The use of a spectral weight with a single peak is motivated by the recent
neutron scattering measurements of the magnetic resonance mode and speculation
that it could be at the origin of d-wave superconductivity in the cuprates.
The results are compared with recent SIS tunneling measurements.\cite{JFZ}

\section {METHOD}

The spectral weight function used in this work is given by
\begin{equation}
\alpha^{2} \; {\rm F}(\omega) \; = \; 
[ \; c_{S} \; + \; c_{D} \; {\rm cos}(2( \phi \; - \phi^{'})) \;]
\; f(\omega) 
\end{equation}
where $f(\omega)$ is a single peaked Lorentzian, a typical example of which is 
depicted in Figure 1 where $E \; = \; \hbar \; \omega$. \\
$\phi$ represents the angular position on a two dimensional Fermi surface. 
This type of interaction which combines an s-wave 
coupling ($c_{S}$) and a d-wave coupling ($c_{D}$) has been studied in the
context of mixed s and d-wave superconducting states \cite{Joynt,Um}. In \cite{Joynt}, the d-wave
phase was preferred for sufficiently small values of $c_{S}/c_{D}$. In 
the present study, this ratio is varied from 0 up to 0.4.

Assuming a strong coupling renormalization $Z(\omega)$ 
that is independent of angle $\phi$ and a 
d-wave superconducting gap
function that can be written as $\Delta(\omega) \; {\rm cos}( 2\phi)$, the
conventional Eliashberg strong coupling equations \cite{Schrieffer} can be written in the form 
\begin{equation}
\omega \; Z(\omega) \; = \; \omega \; - \;
\int_{0}^{\infty} d\omega^{'} \; \int_{0}^{2 \pi} \frac{ d \phi^{'}}{2 \pi} 
\frac{c_{S} \; \omega^{'} \; K_{-}(\omega^{'}, \omega)}
{ \surd \omega^{'2} \; - \; \Delta^{2}(\omega^{'}) \; {\rm cos}^{2}(2 \phi^{'} ) }
\end{equation}
and 
\begin{equation}
\Delta(\omega) \; = \; \frac{1}{Z(\omega)} \; 
\int_{0}^{\infty} d\omega^{'} \; \int_{0}^{2 \pi} \frac{d \phi^{'}}{2 \pi} 
\frac{c_{D} \; \Delta(\omega^{'}) \; {\rm cos}^{2}(2 \phi^{'} )
\; K_{+}(\omega^{'}, \omega)}
{ \surd \omega^{'2} \; - \; \Delta^{2}(\omega^{'}) \; {\rm cos}^{2}(2 \phi^{'} ) }
\end{equation}
where
\begin{equation}
K_{+/-}(\omega^{'}, \omega)= \int_{0}^{\infty} \; d \omega^{''} \; \alpha^{2} F( \omega^{''}) \;
( \; \frac{1}{\omega \; + \; \omega^{'} + \omega{''} \; + \; i \delta}
\; -/+
\;  \frac{1}{\omega \; - \; \omega^{'} - \omega{''} \; + \; i \delta} )
\end{equation}
The density of states is given by
\begin{equation}
{\rm N}(\omega) \; = \; \frac{1}{2 \pi} \; \int_{0}^{2 \pi} \;  d \phi \; 
{\rm Real} \; \frac{\omega}
{\surd( \omega^{2} \; - \; \Delta^{2}(\omega) \;  {\rm cos}^{2}(2 \phi ) }
\end{equation}
and the SIS tunneling conductance curves are calculated from dI/dV where
\begin{equation}
I \; = \; 
\int_{-\infty}^{\infty} d \omega \; N(\omega) \; N(\omega + eV) \; [f(\omega) \; 
- \; f(\omega + eV)] 
\end{equation}

\section{Discussion}

Results for d-wave SIS conductance (dI/dV) curves are presented in figures 3 and 4. 
Tables I and II provide a list of the values of the gap parameter, $\Delta$, the dip position $E_{dip}$
obtained from the SIS conductance curves and the corresponding parameter values for
$c_{S}$, $c_{D}$ and $E_{mode}$. For the results shown in figure 3, the position of the 
peak in the spectral weight function $f(E)$ of equation (1) is kept fixed at $E_{mode}=1.0$, 
the coupling constant $c_{S}$ is increased from 0 to 0.4 and $c_{D}$ is kept fixed at 1.0.
For figure 4, the position of the peak in the spectral weight function $f(E)$ is 
decreased from $E_{mode} = 0.7$ to $0.4$, the coupling constant $c_{S}$ is increased from
0.25 to 0.4 and $c_{D}$ is kept fixed at 1.0.

The results shown in figures 3 and 4 show variations in the calculated position of
the dip in the calculated SIS conductance curves as the ratio $c_{S}/c_{D}$ is changed.
In calculating these curves, the 
experimental measurements \cite{JFZ}, which give values for the ratio
$E_{mode}/\Delta$ ranging from 0.5 (underdoped) up to 2.0 (overdoped), provide
a useful guide for the overall magnitude of $\alpha^{2}F(\omega)$ in equations (2) and (3).
For comparison, for an s-wave superconducting state to yield 
values of $E_{mode}/\Delta$ in the same range as the experiments on the Bi-2212 crystals,
a strong coupling calculation based on a
spectral weight function of the type shown in figure 1 would require a 
$\lambda \; = \; 2 \;\int \alpha^{2}F(E)/E$ of the order of 2.0.
This would result in the type of density of states $N(E)$ curve shown in figure 2.

In the dI/dV curves shown in figures 3 and 4, the energy separation between the 
main conductance peak at $2\Delta$ and the dip at $E_{dip}$, which is listed in
column 3 of Tables I and II, is approximately 
equal to the energy of the mode in the spectral weight function $f(E)$. 
This approximate equality begins to break down somewhat as $c_{S}$ gets larger in both sets of
results. 

As the energy of the peak in $f(E)$, $E_{mode}$, is decreased and $c_{S}$ is increased 
in the calculations leading to figure 4, the superconducting gap decreases from 0.42 to slightly 
less than 0.19 and the dip feature moves away from the main tunneling conductance peak while
weakening in strength.  The energy separation between the conductance peak at $2 \Delta$ and 
the dip at $E_{dip}$ decreases and is close to the mode energy in all four curves of figure 4, 
with the largest deviation occuring for curve D. Furthermore, the ratio of the mode energy to 
the superconducting gap energy ($E_{mode}/ \Delta$ ) ranges from  1.67 up to 2.1.  
The variation for $E_{mode}/ \Delta$ seen in the curves of figure 4 and the 
approximate equality between the mode energy and the conductance peak-dip energy separation
are consistent with the recent tunneling data on the Bi-2212 crystals \cite{JFZ}.

A broadening of the main SIS conductance peak is evident in some of the curves of
Figures 3 and 4. The peak is at its sharpest when $c_{S}$ is largest (curve E of figure 
3 and curve D of figure 4). The decrease in the overall magnitude of the imaginary part of the gap, 
Im$\Delta(\omega)$,  which is shown in figure 5, for curves A and D of figure 4, 
is seen to be at the origin of the sharpening of the main dI/dV peaks in figure 4. Similiar 
behavior is present in the results leading to figure 3.  A sharpening of the main SIS
tunneling conductance peak with decreasing superconducting gap is also observed in
the results of \cite{JFZ}.

Finally, the effect of choosing a more sharply peaked $f(E)$ in the spectral weight function of 
equation (1) is depicted in Figure 6 where the superconducting density of states is labelled as
Curve A and the corresponding SIS conductance as Curve B. The $f(E)$ function for these 
curves is depicted in Figure 7.

\section{Conclusion}

Results are presented for the SIS conductance for a d-wave symmetry superconducting state 
using a conventional Eliashberg strong coupling formalism incorporating the model spectral weight 
of equation (1). The aim is to study the position of a dip feature which is a signature
of the single peak in the spectral weight.

It is possible, using the model spectral weight of equation (1), to arrange for the 
energy difference between the positions of main SIS conductance peak and the dip minimum to 
be comparable to the energy of the spectral weight peak, which is denoted by $E_{mode}$ in this 
work. Recent SIS tunneling measurements \cite{JFZ} have noted this correspondence. The results 
presented here are a consequence of variations in the ratio of the coupling constants,
$c_{S}/c_{D}$, of the model spectral weight function of equation (1) and are
used to model the influence of oxygen doping variations on the
electronic properties of the Bi2212 crystals in the experiment \cite{JFZ}. 
To more fully exploit the recent SIS experimental observations \cite{JFZ},
more sophisticated strong coupling calculations incorporating the tight band structure
appropriate to the cuprates are required in which the 
effect of variations in the chemical potential $\mu$ on the properties of
the spin fluctuation spectral weight and the resulting d-wave superconductivity
can be studied. Of particular interest would be the variation in position
of the dip feature in the density of states and the relationship of this
position to the energy of the $(\pi,\pi)$ peak in the spin fluctuation
susceptibility.

The author wishes to acknowledge useful conversations with J. F. Zasadzinski
and D. Coffey.

%
%

\newpage

\begin{table}
\caption{Figure 3}
\begin{tabular}{ccccccc}
Curve & 2 $\Delta$ & E$_{Dip}$-2 $\Delta$ & E$_{mode}$ & C$_{S}$ & C$_{D}$ & E$_{mode}$/$\Delta$\\
\hline
 A & 0.96 & 0.8 & 1.0  & 0.0 & 1.0 & 2.08 \\
  B& 0.96 & 0.88 & 1.0 & 0.1 & 1.0 & 2.08\\
 C & 1.08 & 0.92 &1.0 & 0.2 & 1.0 & 1.85\\
  D  & 0.72 &1.12 & 1.0 &0.3 & 1.0 & 2.78 \\
  E  & 0.56 & 1.2 & 1.0 & 0.4 & 1.0 & 3.57\\
\end{tabular}
\end{table}

\vspace{1.0in}

\begin{table}
\caption{Figure 4}
\begin{tabular}{ccccccc}
Curve & 2 $\Delta$ & E$_{Dip}$-2 $\Delta$ & E$_{mode}$ & C$_{S}$ & C$_{D}$ & E$_{mode}$/$\Delta$\\
\hline
 A & 0.84 & 0.72 & 0.7  & 0.25 & 1.0 & 1.67 \\
  B& 0.78 & 0.67 & 0.6 & 0.3 & 1.0 & 1.54\\
 C & 0.55 & 0.61 & 0.5 & 0.35 & 1.0 & 1.81\\
  D  & 0.39 & 0.55 & 0.4 &0.4 & 1.0 & 2.1 \\
\end{tabular}
\end{table}

\newpage

%
%

\newpage

\begin{figure}
\begin{center}
\epsfig{file=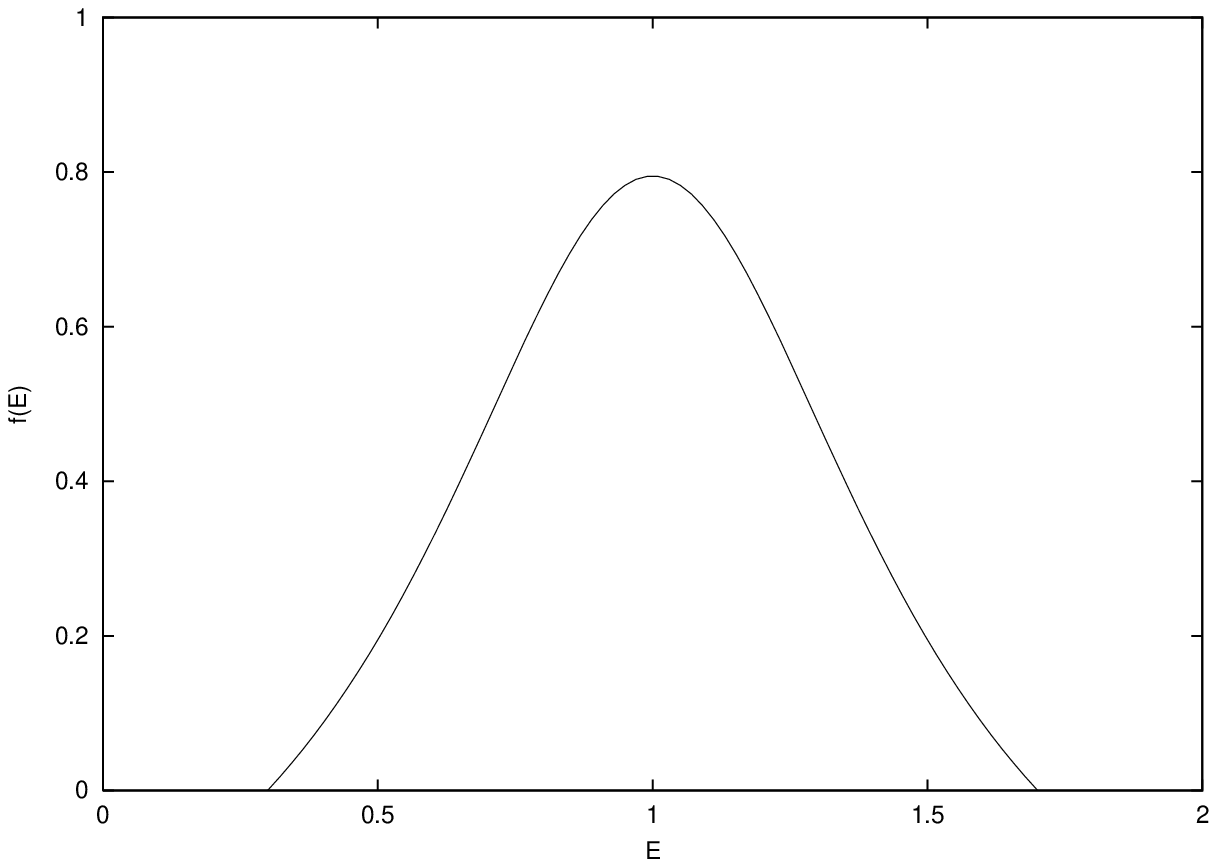,width=4.0in}
\end{center}
\caption{Typical $f(E)$ function used in the $\alpha^{2}F(E)$ of equation (1). The
position of the peak in $f(E)$ is denoted by $E_{mode}$ in the text.}
\end{figure}

\begin{figure}
\begin{center}
\epsfig{file=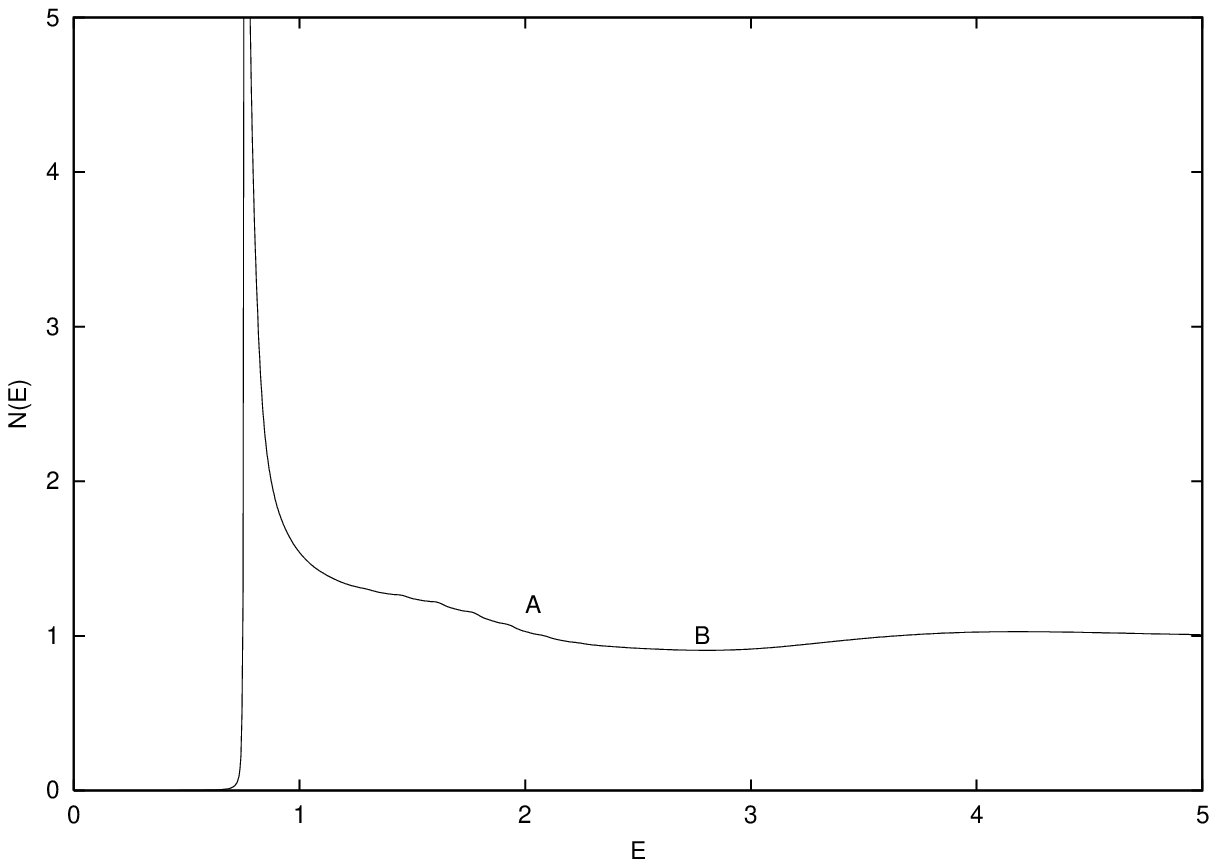,width=4.0in}
\end{center}
\caption{Superconducting density of states N(E) for an s-wave superconductor with
a single peak $\alpha^{2}F(E)$ at $E_{mode}=1.0$ in the strong coupling limit. Point A denotes
the onset of the dip and point B denotes the dip minimum. }
\end{figure}

\newpage

\begin{figure}
\begin{center}
\epsfig{file=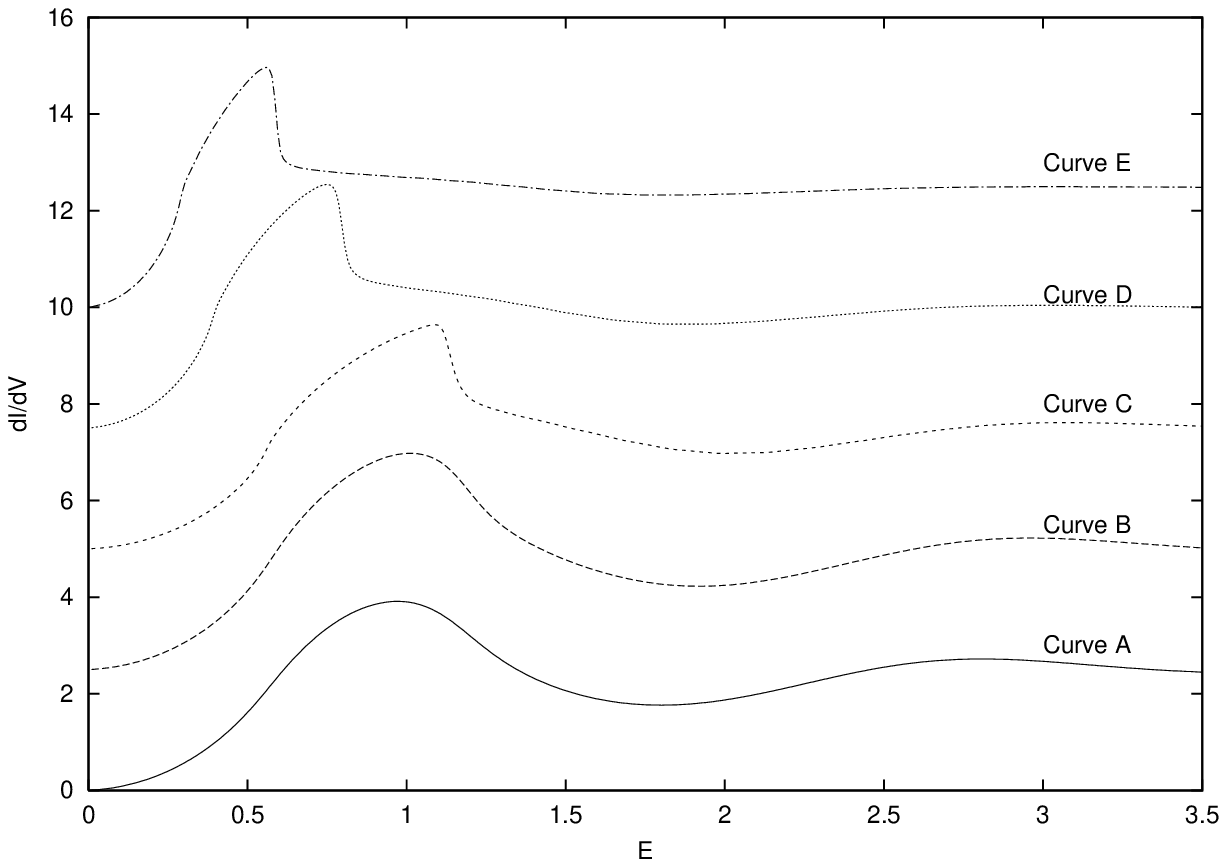,width=6.8in}
\end{center}
\caption{SIS conductance curves for the d-wave superconducting state. The position of
the spectral weight peak is held fixed in these curves. See Table 1.}
\end{figure}

\newpage

\begin{figure}
\begin{center}
\epsfig{file=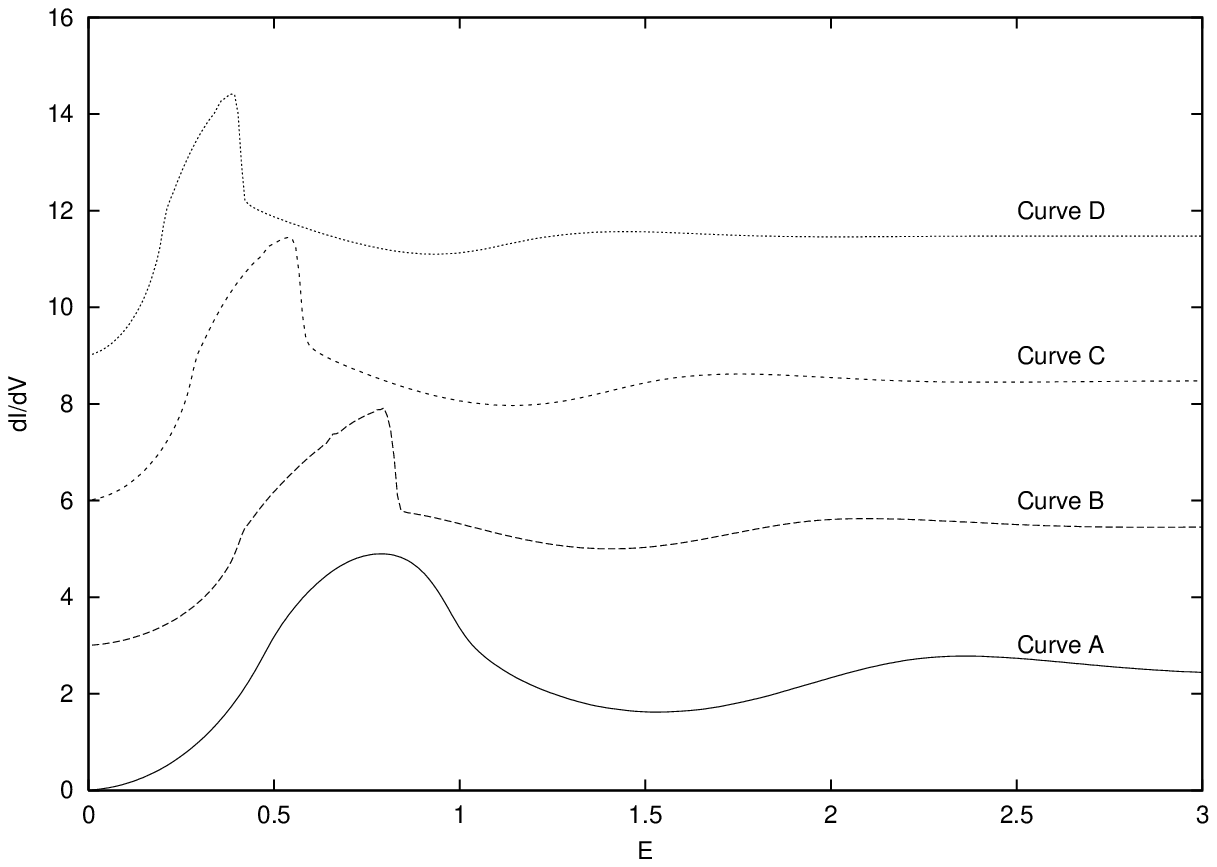,width=6.8in}
\end{center}
\caption{SIS conductance curves for the d-wave superconducting state. The position of
the spectral weight peak is varied in these curves. See Table 2.}
\end{figure}

\newpage

\begin{figure}
\begin{center}
\epsfig{file=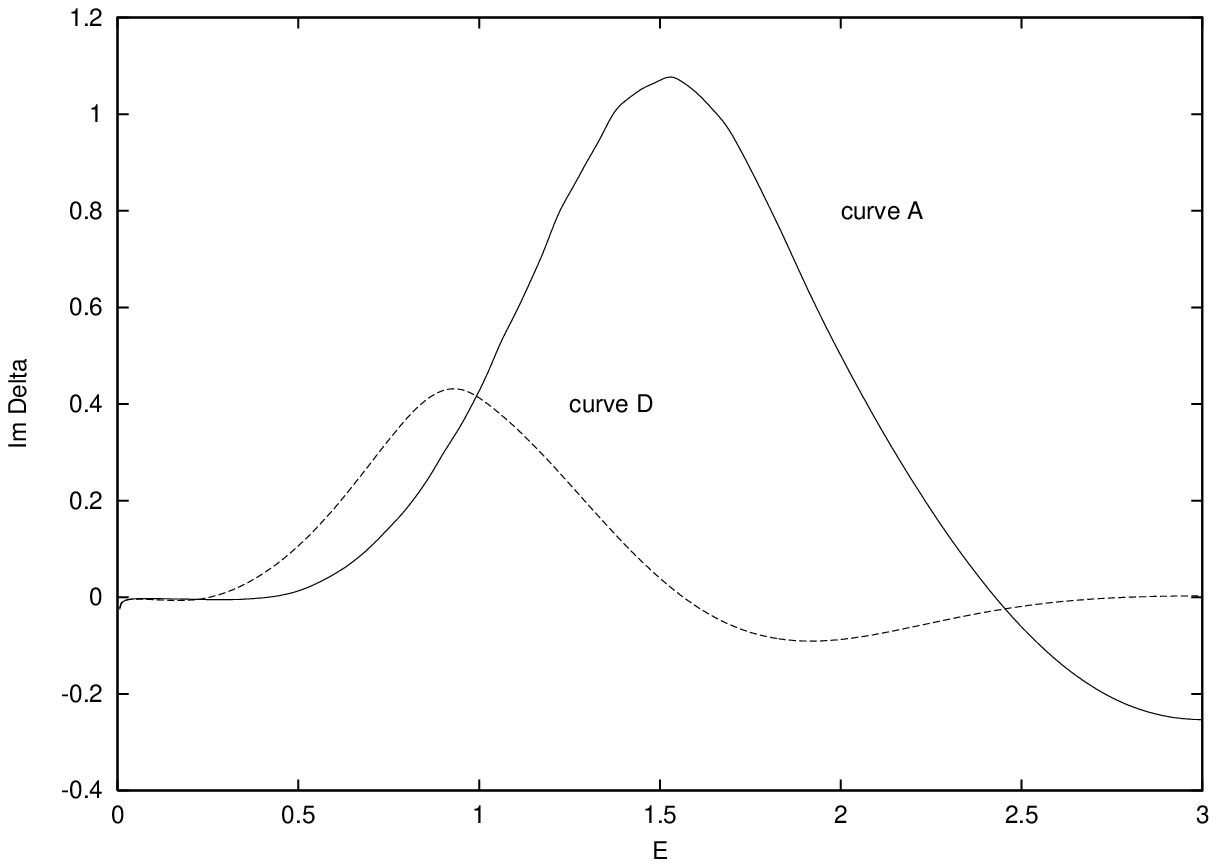,width=6.8in}
\end{center}
\caption{The imaginary part of the superconducting gap for curves A and D of figure 4}
\end{figure}

\newpage

\begin{figure}
\begin{center}
\epsfig{file=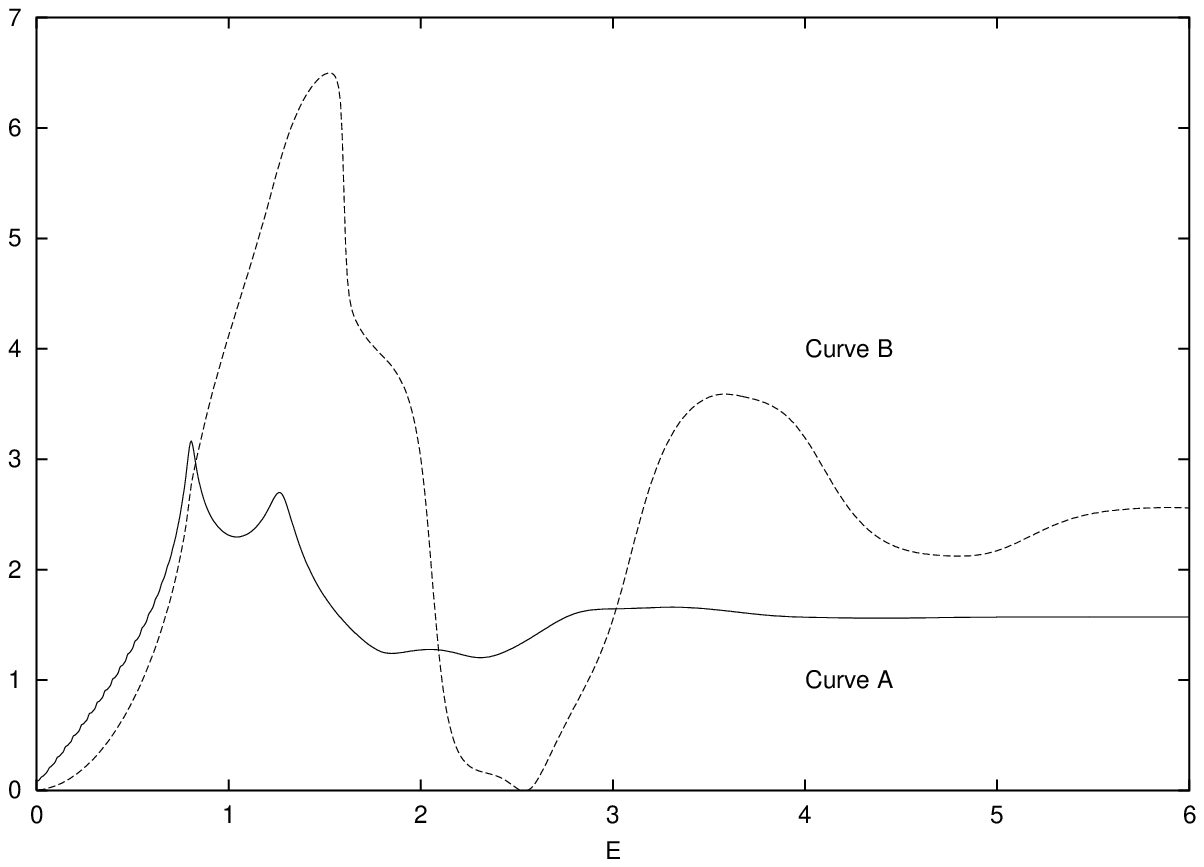,width=4.0in}
\end{center}
\caption{The quasiparticle density of states (Curve A) and the corresponding SIS conductance
(Curve B) for a d-wave symmetry state with a narrow spectral weight peak function $f(E)$
depicted in Figure 7. }
\end{figure}

\begin{figure}
\begin{center}
\epsfig{file=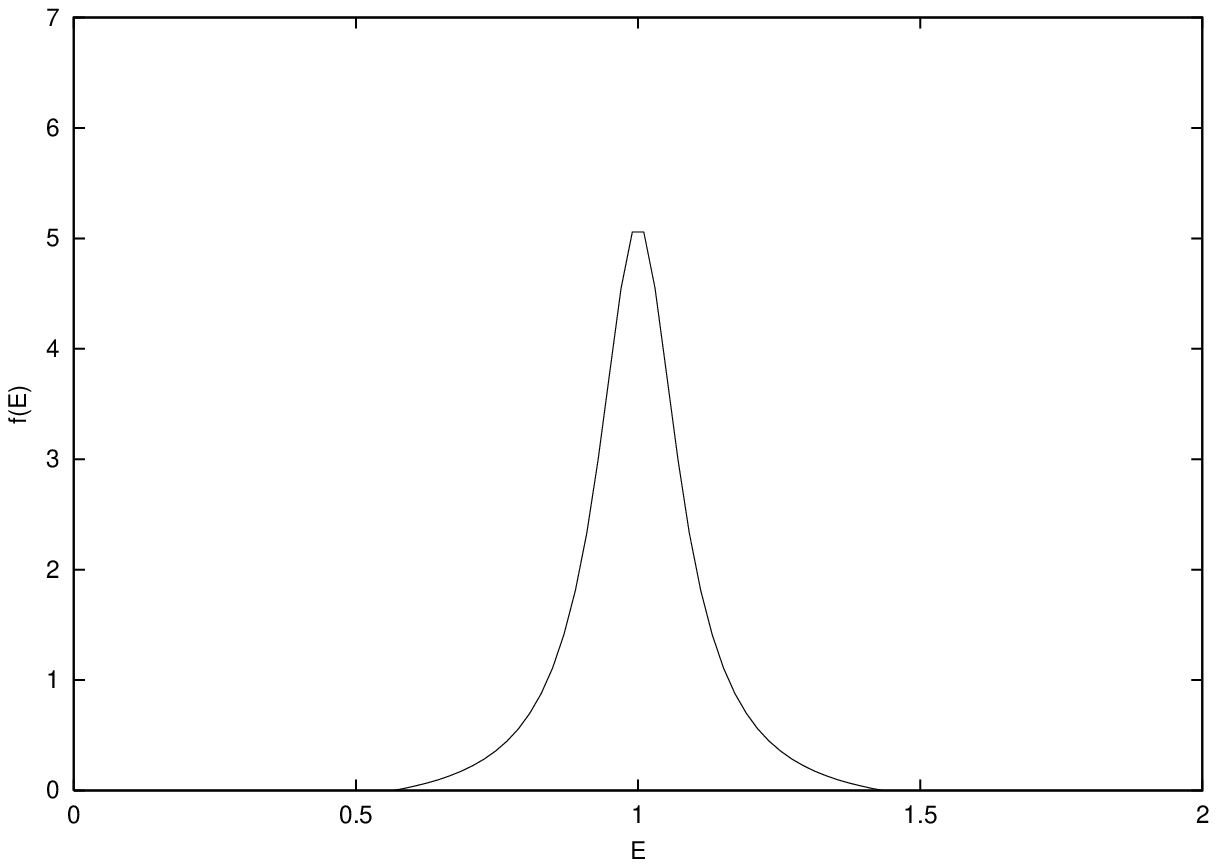,width=4.0in}
\end{center}
\caption{The $f(E)$ function used in the $\alpha^{2}F(E)$ of equation (1) to
generate the results shown in Figure 6.}
\end{figure}


\begin{references}
\bibitem{JFZ} J.F. Zasadzinski, L. Ozyuzer, N. Miyakawa, K.E. Gray, D.G. Hinsk and C. Kendziora,
cond-mat/0102475.
\bibitem{Coffey} D. Coffey and L. Coffey, Phys. Rev. Lett., {\bf 70}, 1529 (1993).
L. Coffey and D. Coffey, Phys. Rev. B. {\bf 48}, 4184 (1993).
\bibitem{Manske} D. Manske, I. Eremin and K.H. Bennemann, Phys. Rev. B, {\bf 63} (2001). 
\bibitem{Schrieffer} J. R. Schrieffer, {\em Theory of Superconductivity}, 3rd Edition,
(Benjamin/Cummings, 1983).
\bibitem{Joynt} K.A. Musaelin, J.Betouras, A.V. Chubukov and R. Joynt,
Phys. Rev. B. {\bf 53}, 3598 (1996).
\bibitem{Um} G.A. Ummarino, R.S. Gonnelli and D. Daghero, cond-mat/0011472
\end{references}
\end{document}